\documentclass[12pt,preprint]{aastex}

\usepackage{graphicx}
\usepackage{epstopdf}
\usepackage{chngpage}

\newcommand{\kms}{km s$^{-1}$}
\newcommand{\ha}{H$\alpha$}
\newcommand{\solar}{\ifmmode_{\sun}\else$_{\sun}$\fi}

\newcommand{\HI}{H$\,${\sc i}}

\begin{document}

\title{A study of two dwarf irregular galaxies with asymmetrical star formation distributions}

\author{
Deidre A.\ Hunter\altaffilmark{1}, 
Samavarti Gallardo\altaffilmark{1,2}, 
Hong-Xin Zhang\altaffilmark{3},
Angela Adamo\altaffilmark{4},
{\bf David} O.\ Cook\altaffilmark{5},
Se-Heon Oh\altaffilmark{6},
Bruce G.\ Elmegreen\altaffilmark{7},
Hwihyun Kim\altaffilmark{8},
Lauren Kahre\altaffilmark{9},
Leonardo Ubeda\altaffilmark{10},
Stacey N.\ Bright\altaffilmark{10},
Jenna E.\ Ryon\altaffilmark{10},
Michele Fumagalli\altaffilmark{11},
Elena Sacchi\altaffilmark{12,13},
R.\ C. Kennicutt \altaffilmark{14},
Monica Tosi\altaffilmark{13},
Daniel A.\ Dale\altaffilmark{15},
Michele Cignoni\altaffilmark{16,17},
Matteo Messa\altaffilmark{4},
Eva K.\ Grebel\altaffilmark{18},
Dimitrios M.\ Gouliermis\altaffilmark{19,20},
Elena Sabbi\altaffilmark{10},
Kathryn Grasha\altaffilmark{21},
John S.\ Gallagher, III\altaffilmark{22},
Daniela Calzetti\altaffilmark{21},
Janice C.\ Lee\altaffilmark{23}
}

\altaffiltext{1}{Lowell Observatory, 1400 West Mars Hill Road, Flagstaff, Arizona 86001 USA}
\altaffiltext{2}{Northern Arizona University, Flagstaff, Arizona 86011 USA}
\altaffiltext{3}{Institute of Astrophysics, Pontificia Universidad Cat\'{o}lica de Chile, Av. Vicuna Mackenna 4860, 7820436 Macul, Santiago, Chile}
\altaffiltext{4}{Department of Astronomy, The Oskar Klein Centre, Stockholm University, Stockholm, Sweden}
\altaffiltext{5}{Department of Astronomy and Astrophysics, California Institute of Technology, Pasadena, California 91125 USA}
\altaffiltext{6}{Korea Astronomy and Space Science Institute, Daedeokdae-ro 776, Yuseong-gu, Daejeon 34055 Republic of Korea}
\altaffiltext{7}{IBM Research Division, T.\ J.\ Watson Research Center, Yorktown Heights, New York 10598 USA}
\altaffiltext{8}{Gemini Observatory, Casilla 603, La Serena, Chile}
\altaffiltext{9}{Department of Astronomy, New Mexico State University, Las Cruces, New Mexico USA}
\altaffiltext{10}{Space Telescope Science Institute, 3700 San Martin Drive, Baltimore, Maryland 21218 USA}
\altaffiltext{11}{Institute for Computational Cosmology and Centre for Extragalactic Astronomy, Durham University, Durham, United Kingdom}
\altaffiltext{12}{Department of Physics and Astronomy, Bologna University, Bologna, Italy}
\altaffiltext{13}{INAF -- Osservatorio Astronomico di Bologna, Bologna, Italy}
\altaffiltext{14}{Institute of Astronomy, University of Cambridge, Cambridge, United Kingdom}
\altaffiltext{15}{Deptartment of Physics and Astronomy, University of Wyoming, Laramie, Wyoming USA}
\altaffiltext{16}{Department of Physics, University of Pisa, Largo B. Pontecorvo 3, 56127, Pisa, Italy }
\altaffiltext{17}{INFN, Largo B. Pontecorvo 3, 56127, Pisa, Italy}
\altaffiltext{18}{Astronomisches Rechen-Institut, Zentrum f\"ur Astronomie der Universit\"at Heidelberg, M\"onchhofstr.\ 12--14, 69120 Heidelberg, Germany}
\altaffiltext{19}{Zentrum f\"ur Astronomie der Universit\"at Heidelberg, Institut f\"ur Theoretische Astrophysik, Albert-Ueberle-Str.\,2, 69120 Heidelberg, Germany}
\altaffiltext{20}{Max Planck Institute for Astronomy,  K\"{o}nigstuhl\,17, 69117 Heidelberg, Germany}
\altaffiltext{21}{Department of Astronomy, University of Massachusetts -- Amherst, Amherst, Massachusets 01003 USA}
\altaffiltext{22}{Department of Astronomy, University of Wisconsin--Madison, Madison, Wisconsin USA}
\altaffiltext{23}{IPAC, Division of Physics, Mathematics, and Astronomy, California Institute of Technology, 1200 E.\ California Blvd., Pasadena, California 91125 USA}

\begin{abstract}
Two dwarf irregular galaxies DDO 187 and NGC 3738 exhibit a striking pattern of star formation:
intense star formation is taking place in a large region occupying roughly 
half of the inner part of the optical galaxy. 
We use data on the \HI\ distribution and kinematics and stellar images and colors to examine the properties of the environment
in the high star formation rate (HSF) halves of the galaxies in comparison with the low star formation rate (LSF) halves.
We find that the pressure and gas density are higher on the HSF sides by 30-70\%.
In addition we find in both galaxies that the \HI\ velocity fields exhibit significant deviations from ordered rotation
and there are large regions of high velocity dispersion and multiple velocity components
in the gas beyond the inner regions of the galaxies.
The conditions in the HSF regions are likely the result of large-scale external processes
affecting the internal environment of the galaxies and enabling the current star formation there.
\end{abstract}


\section{Introduction} \label{sec-intro}

Dwarf irregular (dIrr) galaxies are defined by their irregular morphology. In many relatively isolated dIrrs, star-forming regions
are scattered seemingly randomly across the disk, in hierarchical clusterings \citep[for example,][]{ee98a},
or as a second generation of stars triggered by the preceding generation \citep[for example,][]{dopita85,ee98b}.
However, what powers star formation and patterns of star formation in dIrrs is not known since the standard models
for driving molecular cloud formation do not always apply in dwarfs \citep[for example,][]{eh15}.

Local environmental conditions within galaxies have been shown to have consequences on the star formation products. 
For example, \citet{adamo-m83} found that the fraction of star formation resulting in bound clusters, $\Gamma$, 
decreases by factors of a few from the center of M83 to the outer disk and
varies from region to region within the galaxy with the local star formation rate (SFR) density. 
In addition large-scale yet local processes have been found to affect the location of some star-forming regions:
streaming of gas around bar potentials, for example, can cause a pile up of gas at the ends of the bars which creates large star forming
regions \citep{ee80}.

Here we investigate the properties of two nearby dIrrs, DDO 187 and NGC 3738, whose recent star formation activity
is located in one half of the inner part of the optical galaxy. These regions stand out particularly
in contrast to the lower star formation activity in the other half of the galaxy.
This study is an attempt to understand what may be driving star formation in these particular regions today. 
We examine the local environments within the high and low star formation halves and the larger-scale galactic environment
in which these regions are located.

In Section \ref{sec-data} we give a description of the galaxies, the data we have at our disposal,
how we characterize the galactic environments of the star-forming regions, and how we define those regions.
Our observations are described in Section \ref{sec-results}, including a comparison of the galactic characteristics
in the high SFR side and the lower SFR side of the galaxies, 
ages of compact star clusters, and gas kinematics.
In Section \ref{sec-discuss} we summarize the pertinent observations, draw a comparison of DDO 187 and NGC 3738
with IC 10, a dIrr with a tidal tail, and with tadpole galaxies, and speculate on a plausible overall picture of what has happened in these systems.
Section \ref{sec-summary} is a reprise of the highlights.

\section{The Data} \label{sec-data}

\subsection{Galaxy sample}

Some basic properties of DDO 187 and NGC 3738 are given in Table \ref{tab-sample}.
DDO 187 and NGC 3738 are part of the
LITTLE THINGS\footnote[24]{ Funded in part by the
National Science Foundation through grants AST-0707563, AST-0707426, AST-0707468, and
AST-0707835 to US-based LITTLE THINGS team members and with generous technical and logistical support from the
National Radio Astronomy Observatory.} \citep[Local Irregulars That Trace Luminosity
Extremes, The \HI\ Nearby Galaxy Survey;][]{lt} sample.
LITTLE THINGS is a multi-wavelength survey of 37 dIrr galaxies and 4 Blue Compact Dwarfs (BCD)
aimed at understanding what drives star formation in tiny systems. 
The LITTLE THINGS galaxies were chosen to be nearby ($\leq$10.3 Mpc), contain gas so they could be forming stars,
and cover a large range in dwarf galactic properties.

\begin{deluxetable}{lcccccccc}
\tabletypesize{\scriptsize}
\tablecolumns{9}
\tablewidth{450pt}
\tablecaption{The Galaxy Sample \label{tab-sample}}
\tablehead{
\colhead{} & \colhead{} & \colhead{D} & \colhead{} & \colhead{M$_V$}
& \colhead{\tablenotemark{c}$R_D$} 
& \colhead{\tablenotemark{d}log SFR$_D^{FUV}$}  
& \colhead{} & \colhead{} \\
\colhead{Galaxy}  &  \colhead{Other names\tablenotemark{a}}
& \colhead{(Mpc)} & \colhead{Ref\tablenotemark{b}} & \colhead{(mag)}
& \colhead{(kpc)} 
& \colhead{(M\solar\ yr$^{-1}$ kpc$^{-2}$)} 
& \colhead{$12+\log {\rm (O/H)}$} & \colhead{Ref\tablenotemark{e}}
}
\startdata
DDO 187   & UGC 9128 & 2.2   & 1  & -12.7$\pm$0.014 & $0.18 \pm 0.01$ 
& $-1.98 \pm 0.01$  & 7.75$\pm$0.05 & 3\\
NGC 3738 & UGC 6565 & 4.9 & 2  & -17.1$\pm$0.001 & $0.78 \pm 0.01$ 
& $-1.53 \pm 0.01$ & 8.04$\pm$0.06 & 3 \\
\enddata
\tablenotetext{a}{Selected alternate identifications obtained from NED.}
\tablenotetext{b}{Reference for the distance to the galaxy.}
\tablenotetext{c}{$R_D$ is the disk scale length measured from $V$-band images.
From \citet{HE06} revised to the distance adopted here.}
\tablenotetext{d}{
SFR$_D^{FUV}$ is the integrated star formation rate determined from {\it GALEX} FUV fluxes
\citep[][with an update of the  {\it GALEX} FUV photometry to the GR4/GR5 pipeline reduction]{HEL10},
normalized to the area $\pi R_D^2$, where $R_D$ is the disk scale length.}
\tablenotetext{e}{Reference for the oxygen abundance of the galaxy.}
\tablerefs{
1 -- \citet{dist-d187};
2 -- \citet{dist-n3738};
3 -- \citet{oh-d187}
}
\end{deluxetable}

NGC 3738 is also part of the 
{\it Hubble Space Telescope} ({\it HST}) Legacy Extragalactic UV Survey \citep[LEGUS,][]{legus}.
LEGUS is an {\it HST} Cycle 21 Treasury survey aimed at exploring star formation from scales of
individual stars to kpc-size structures with multi-band imaging on 50 galaxies within 12 Mpc.
The galaxies span the range of star-forming disk galaxies, including dIrrs.

\subsection{Galactic environments}

\begin{figure}
\epsscale{0.8}
\vskip -1.6truein
\plotone{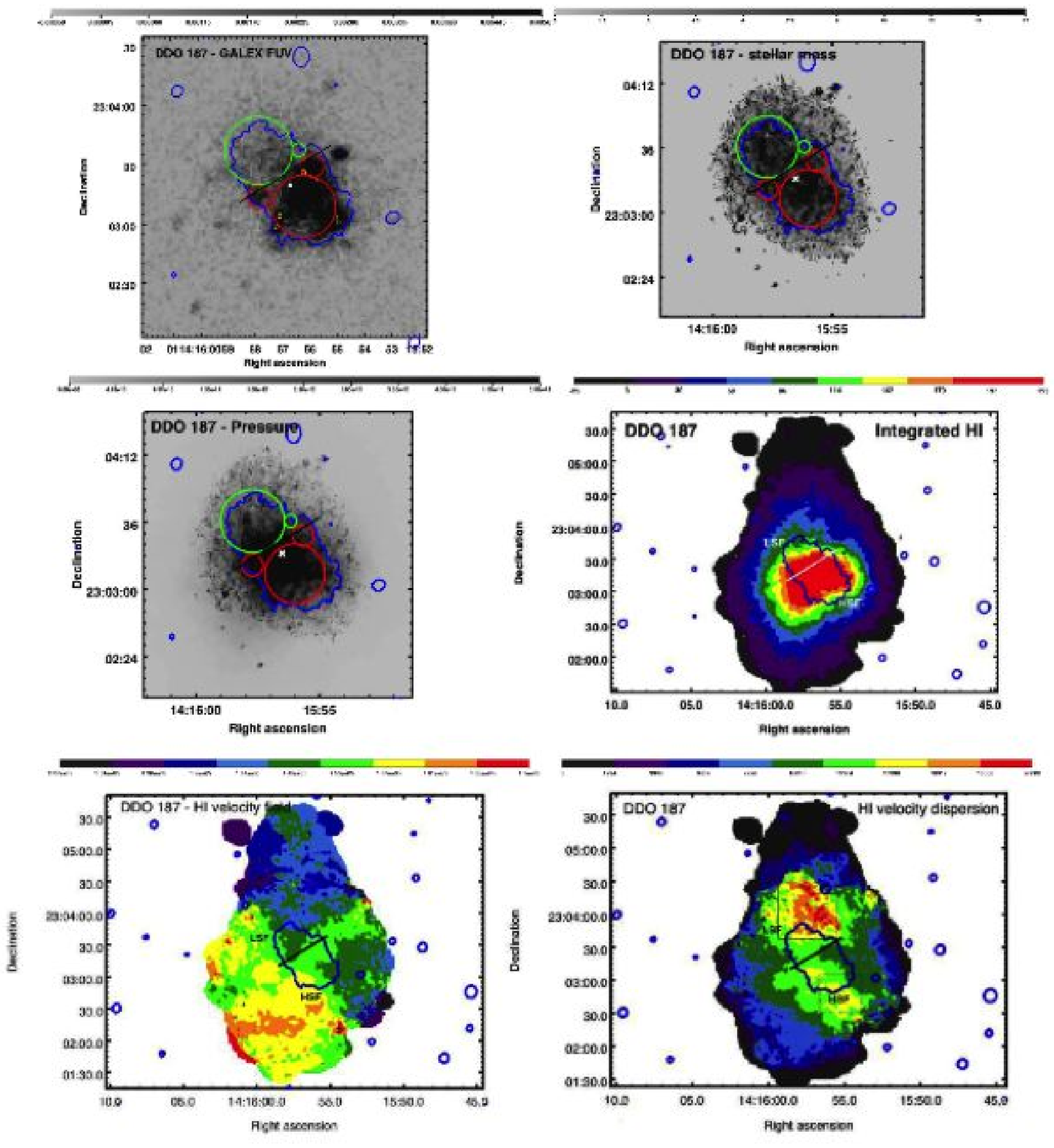}
\caption{\small
DDO 187 images:
{\it Top left:} {\it GALEX} FUV-band image. 
Units of colorbar are counts s$^{-1}$, which can be converted to erg s$^{-1}$ cm$^{-2}$ \AA$^{-1}$ by multiplying by $1.4\times10^{-15}$. 
The bright half of the galaxy to the SW
is referred to as the HSF half and the fainter half of the galaxy to the NE is referred
to as the LSF half. The black line marks the dividing line between the two halves.
The blue contour denotes the 23.6 mag arcsec$^{-2}$ isophote on the $V$-band image and was chosen
to encompass the part of the optical galaxy considered here.
The red circles added together are used to determine the characteristics of the HSF half and the green
circles added together are used to determine the characteristics of the LSF half. 
The white ``x'' marks the optical center of the galaxy \citep{HE06}.
The small yellow circles mark the location of compact star clusters.
{\it Tob right:} Stellar mass surface density map.
Units of colorbar are M\solar\ pc$^{-2}$.
{\it Middel left:} Pressure map. 
Units of colorbar are g s$^{-2}$ cm$^{-1}$.
{\it Middle right:} Integrated \HI\ map.
The units of the colorbar are Jy beam$^{-1}$ m s$^{-1}$; multiply by $8.13\times10^{18}$ to obtain column density in units of atoms cm$^{-2}$.
{\it Bottom left:} \HI\ intensity-weighted velocity field.
Units of the color bar are m s$^{-1}$.
{\it Bottom right:} \HI\ intensity-weighted velocity dispersion map.
Units of the colorbar are m s$^{-1}$.
}
\label{fig-d187}
\end{figure}

\begin{figure}
\epsscale{0.8}
\vskip -1.6truein
\plotone{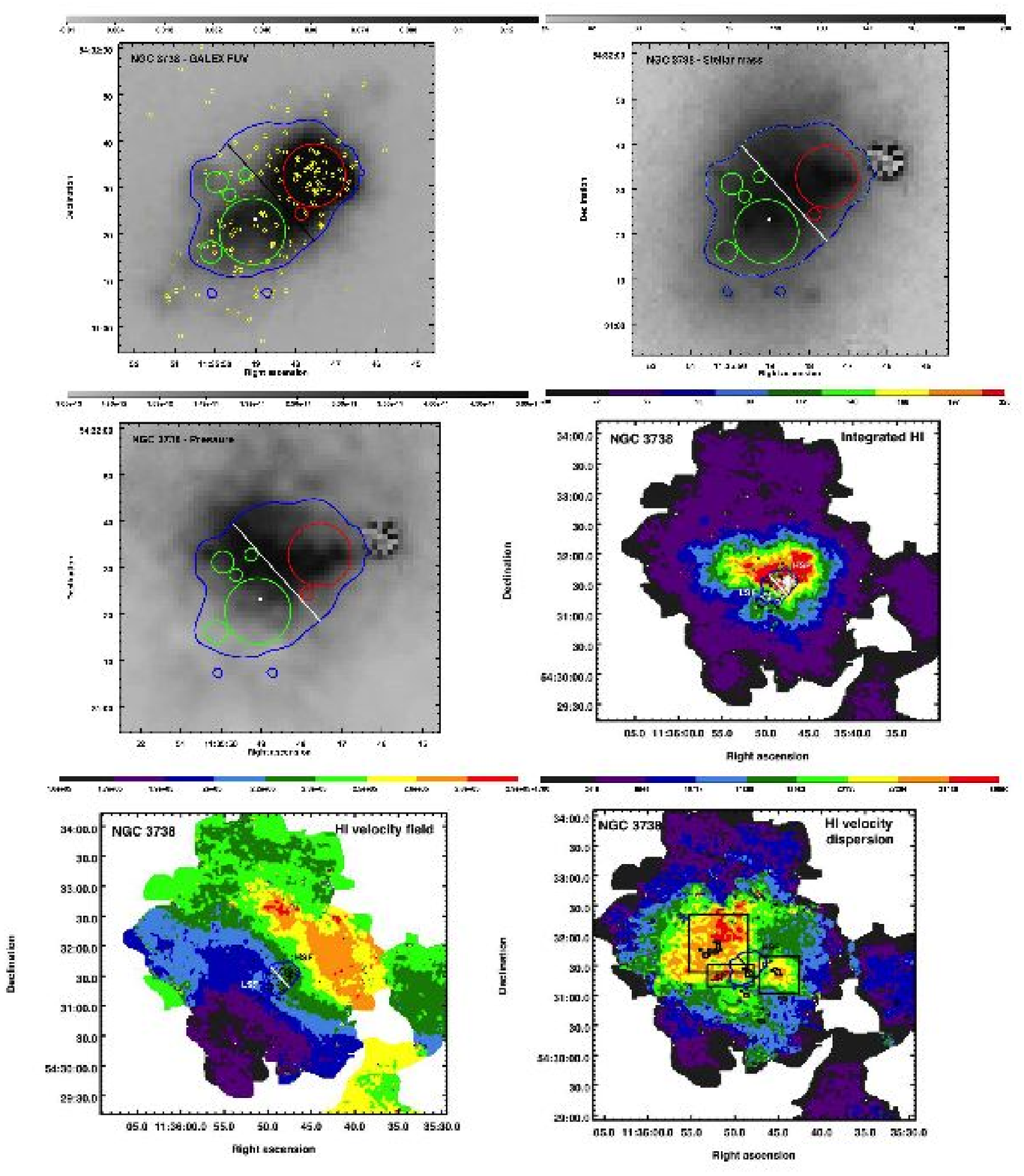}
\vskip -0.5truein
\caption{\small
NGC 3738 images:
{\it Top left:} {\it GALEX} FUV-band image. 
Units of colorbar are counts s$^{-1}$, which can be converted to erg s$^{-1}$ cm$^{-2}$ \AA$^{-1}$ by multiplying by $1.4\times10^{-15}$. 
The bright half of the galaxy to the SW
is referred to as the HSF half and the fainter half of the galaxy to the NE is referred
to as the LSF half. The black line marks the dividing line between the two halves.
The blue contour denotes the 20.6 mag arcsec$^{-2}$ isophote on the $V$-band image and was chosen
to encompass the part of the optical galaxy considered here.
The red circles added together are used to determine the characteristics of the HSF half and the green
circles added together are used to determine the characteristics of the LSF half. 
The white ``x'' marks the optical center of the galaxy \citep{HE06}.
The small yellow circles mark the location of compact star clusters.
{\it Tob right:} Stellar mass surface density map.
Units of colorbar are M\solar\ pc$^{-2}$.
{\it Middel left:} Pressure map. 
Units of colorbar are g s$^{-2}$ cm$^{-1}$.
{\it Middle right:} Integrated \HI\ map.
The small circles mark the location of compact star clusters; white denotes clusters with ages less than or equal to 30 Myr, 
black circles are for clusters with ages between 30 Myr and 100 Myr, and green circles are clusters with ages greater than 100 Myr.
The units of the colorbar are Jy beam$^{-1}$ m s$^{-1}$; multiply by $1.09\times10^{19}$ to obtain column density in units of atoms cm$^{-2}$.
{\it Bottom left:} \HI\ intensity-weighted velocity field.
The black dots are compact star clusters.
Units of the color bar are m s$^{-1}$.
{\it Bottom right:} \HI\ intensity-weighted velocity dispersion map.
The small black contours are the integrated \HI\  with strong non-circular motions from \citet{oh15}.
``Strong'' here means that the intensity of the \HI\ in this velocity component is stronger than the
signal in the gas undergoing ordered rotation. 
The peculiar velocity maps were made from the robust-weighted data cube and have a beam size of 6.26$\times$5.51\arcsec.
Column densities of the gas engaged in non-circular motions are of order $10^{21}$ atoms cm$^{-2}$.
The black boxes outline grids of velocity profiles discussed in Section 3.5.
Units of the colorbar are m s$^{-1}$.
}
\label{fig-n3738}
\end{figure}

To characterize the galactic environments within DDO 187 and NGC 3738 we used LITTLE THINGS data.
The LITTLE THINGS data sets include 
\HI-line maps obtained with the Karl G.\ Jansky Very Large Array (VLA\footnote[25]{
The National Radio Astronomy Observatory is a facility of the National Science Foundation operated under 
cooperative agreement by Associated Universities, Inc.}), and we used the naturally-weighted maps here.
In natural-weighting the visibilities have the same weights resulting in somewhat higher sensitivity but poorer spatial resolution compared
to other weighting schemes.
For DDO 187, the channel separation was 1.3 \kms, the naturally-weighted beam FWHM was 
12\farcs4$\times$11\farcs0, and the single channel rms was 0.54 Jy beam$^{-1}$.
For NGC 3738, the channel separation was 2.6 \kms, the beam FWHM was 
13\farcs0$\times$7\farcs8, and the single channel rms was 0.46 Jy beam$^{-1}$.
$\Sigma_{gas}$ is the \HI\ surface density multiplied by 1.34 to account for Helium \citep{helium}. 
We also used the intensity-weighted velocity field (moment 1) and velocity dispersion (moment 2) maps.
The \HI\ maps are shown in Figures \ref{fig-d187} and \ref{fig-n3738}.

In addition, we used $B$ and $V$ images obtained with the Hall 1.1-meter telescope at Lowell Observatory to determine the 
stellar mass density on a pixel by pixel basis. 
We used the $B-V$ color to determine the mass-to-light ratio using an empirical relationship
determined from spectral energy distribution (SED) fitting to LITTLE THINGS annular surface photometry \citep{ml}, 
and, with the $V$-band luminosity $L_V$, determined the stellar mass in each pixel.
According to \citet{ml17}, $B$ and $V$ are the best passbands to use for estimating stellar masses from
a single color for dwarf galaxies. 
Adequate near infrared images are not available from {\it WISE} or {\it Spitzer}.
From \citet{ml17} 
we can expect that these stellar masses are good to a factor of two and we adopt an uncertainty of 0.3 dex in the
stellar mass surface densities.

Together the gas and stellar mass maps were used to produce maps of the hydrostatic mid-plane pressure from:
$$ P = 2.934\times10^{-55} \times \Sigma_{gas}\times(\Sigma_{gas} + (\sigma_g/\sigma_*)\times\Sigma_*)  ~~{\rm [g/(s^2 cm)],} $$
where $\Sigma$ is a surface density and $\sigma$ is a velocity dispersion \citep{bruce89}.
The stellar velocity dispersion, $\sigma_*$, was estimated using $\log \sigma_* = -0.15M_B - 1.27$ from \citet{swaters}
where $M_B$ is the integrated $B$-band absolute magnitude of the galaxy.
$\Sigma_{gas}$ is determined from \HI$+$He since the molecular content has not been mapped in these galaxies.
CO has been detected in NGC 3738 at a surface density of 2.7 M\solar\ pc$^{-2}$ in a single pointing 
\citep{youngco,kennicuttco}, but generally \HI\ dominates the apparent gas mass density.
Also, the emphasis here is on the material that is available to become molecular on a larger spatial scale.
We, therefore, have maps of the pressure, gas kinematics, atomic gas mass surface density, and stellar mass surface density
with which to characterize the environment in which recent star formation has taken place.
The stellar mass density and pressure maps are shown in Figures \ref{fig-d187} and \ref{fig-n3738}.

We estimate the uncertainties in $\Sigma_{HI}$ from the \HI\ data cube channel rms given by \citet{lt} and 
assume the number of channels contributing to each pixel in the integrated moment zero map is
a typical FWHM divided by the channel width. Then, knowing the number of pixels summed for each region,
we derive the uncertainty in the \HI\ flux. For NGC 3738 the uncertainties are 4\% and 5\% of the flux in the HSF and LSF
regions, respectively. For DDO 187 the uncertainties are 12\% and 19\%.
The uncertainty in the pressure is determined from the fact that the pressure
is dominated by $\Sigma_{HI}^2$.

\subsection{Star clusters}

The LEGUS team identified compact star clusters in the LEGUS sample of galaxies \citep{leguscl,cook18},
including NGC 3738, and we use the LEGUS catalog to examine the star clusters in each half of the galaxy. 
The clusters are compact and centrally-concentrated sources (class 1 or 2 objects), which could be
gravitationally bound systems,
or those with asymmetric profiles and multiple peaks on top of diffuse underlying wings (class 3) \citep{leguscl}. 
We did not include objects 
whose authenticity as a cluster was not verified because of their faintness (Class 0) 
or which are likely non-cluster contaminants (Class 4) \citep[see][for more details]{grasha15}.
The catalog contains the Concentration Index (CI), 
defined as the magnitude in the F606W filter within an aperture of radius 1 pixel minus the magnitude measured
within an aperture of radius 3 pixels, which is 2.8 pc at NGC 3738.
The catalog also contains ages and masses of the clusters determined from SED fitting to multiple {\it HST} passbands. 
Several internal reddening curves were used, and here we adopted the catalogues
in which the photometry was fit for internal extinction using the curve of \citet{starburstext}.
See \citet{leguscl} for details.

\begin{figure}[t!]
\epsscale{0.7}
\vskip -1truein
\plotone{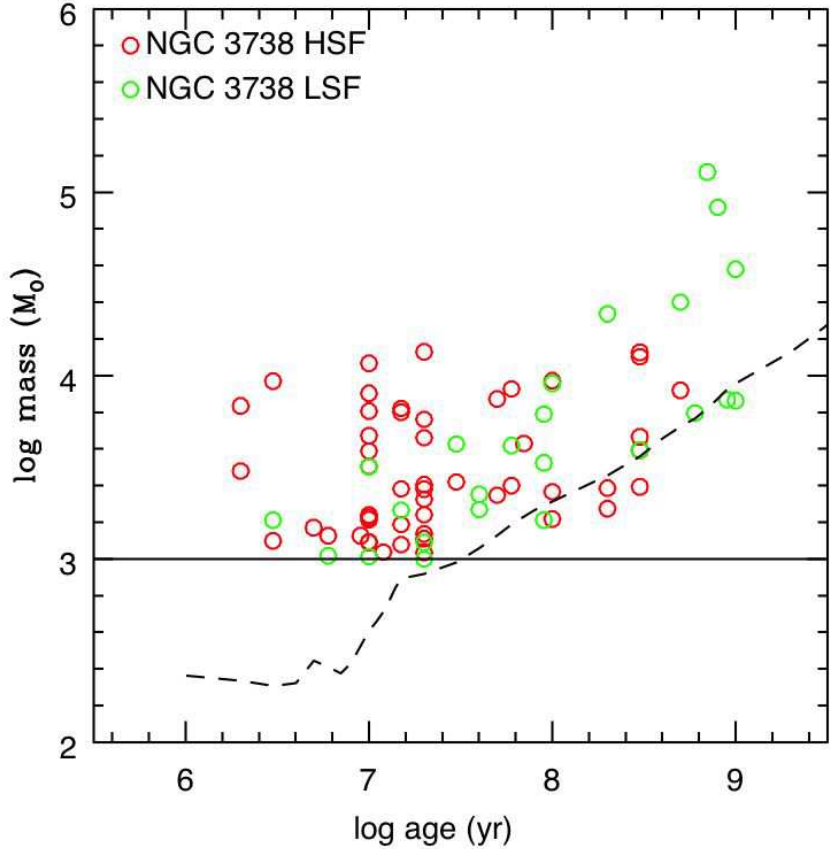}
\vskip -0.25truein
\caption{
Mass versus age for the 75 star clusters in the two regions of NGC 3738 studied here.
Clusters were eliminated on the basis of class (0 and 4)
and number of filters with observations ($<$4 filters). 
The horizontal solid black line marks the cluster mass cut (mass $<$1000 M\solar).
The slanted dashed line marks the catalog limits for visual inspection of the clusters defined as
$M_{F606W}=-6$ for NGC 3738, a conservative estimate of completeness limit \citep{leguscl}.
The concern for incompleteness is at ages older than 35 Myr and extends to higher mass with increasing age.
}
\label{fig-clmassvsage}
\end{figure}

In order to use only the clusters in the catalog whose properties were determined with confidence and whose numbers were fairly complete, 
we eliminated clusters with imaging in less than 4 filters, with masses less than 1000 M\solar, and with undetermined ages.
This left 182 clusters. 
Of these, 75 are in the region we examine here.
The LEGUS team carried out tests on the incompleteness of the cluster catalog of NGC 628 at 9.9 Mpc \citep{leguscl}
and M51 at 7.7 Mpc \citep{legusm51} by counting the recovery of synthetic sources. 
Completeness in the F555W image of NGC 628 is 90\% at about 25.0 mag and of M51 at about 23.7 mag.
The LEGUS cluster catalogues also used an absolute magnitude limit of $-6$ for visual inspection of the clusters \citep{leguscl}.
In NGC 628 and M51 the magnitude cut is brighter than the 90\% completeness limit. 
The completeness limits in other galaxies are expected to be similar.
In Figure \ref{fig-clmassvsage} we plot the mass versus age for the 75 clusters in our regions along with a line 
that shows this magnitude limit projected on the mass-age plane. This represents a conservative estimate of the
completeness of the cluster catalog. 
The concern for incompleteness is at ages older than 35 Myr and increases in mass with age, but
the completeness in both the HSF and LSF regions is similar.
See \citet{clpaper}  for further discussion of the properties of these clusters as well as OB associations and OB stars in 
NGC 3738.

DDO 187 is not part of the LEGUS sample. However, there are two {\it HST} images of this galaxy in the F606W and
F814W filters in the archives.
We used the {\it HST} images and the LEGUS pipeline to identify and verify candidate star clusters in DDO 187.
Thirty-five candidate clusters were classified by eye, and 6 are class 1, 2, or 3.
The positions of these six clusters are marked in Figure \ref{fig-d187}. The CI parameter was measured on the F606W filter. 

We also have $UBVI$ ground-based images obtained under excellent seeing
with Lowell Observatory's 4.3-m Discovery Channel Telescope (DCT) and the Large Monolithic Imager,
and we found it more useful to use these multi-wavelength data to determine ages and masses of the clusters \citep{zhang12}.
On the DCT images the seeing was 0.65\arcsec, 1.2\arcsec, 0.65\arcsec, 0.94\arcsec, for $UBVI$, respectively.
We used Starburst99 models \citep{starburst99} with the Geneva high mass-loss stellar evolutionary tracks \citep{meynet94}, 
a Kroupa stellar initial mass function from 0.1 to 100 M\solar\  \citep{kroupaimf},
and the \citet{starburstext} extinction curve
in order to treat the DDO 187 clusters similarly to the way the LEGUS team fit the clusters of NGC 3738.
The photometry and SED fitting for the 6 clusters in DDO 187 are given in Tables \ref{tab-d187cl} and \ref{tab-d187sed}.

\begin{deluxetable}{lccccccc}
\tabletypesize{\scriptsize}
\tablecolumns{8}
\tablewidth{360pt}
\tablecaption{The Clusters in DDO 187 \label{tab-d187cl}}
\tablehead{
\colhead{ID} & \colhead{RA (h m s)} & \colhead{Dec (\arcdeg\ \arcmin\ \arcsec)} & \colhead{CI\tablenotemark{a}} 
& \colhead{$V$} & \colhead{$B-V$} & \colhead{$U-B$} & \colhead{$V-I$} 
}
\startdata    
1 & 14 15 56.17 & 23 03 25.5 & 1.5 & 21.83$\pm$0.04 & 1.05$\pm$0.08 & -0.61$\pm$0.12 &   0.19$\pm$0.06 \\
2 & 14 15 55.67 & 23 03 21.5 & 1.3 & 22.47$\pm$0.06 & 1.33$\pm$0.14 & -0.65$\pm$0.20 &   0.64$\pm$0.08 \\
3 & 14 15 57.09 & 23 03 03.5 & 1.5 & 22.25$\pm$0.04 & 0.71$\pm$0.05 & -1.40$\pm$0.04 &  -0.46$\pm$0.11 \\
4 & 14 15 55.03 & 23 03 02.7 & 1.5 & 21.92$\pm$0.03 & 0.51$\pm$0.04 & -1.23$\pm$0.04 &  -0.24$\pm$0.07 \\
5 & 14 15 54.95 & 23 03 00.1 & 1.4 & 21.33$\pm$0.02 & 0.73$\pm$0.03 & -1.35$\pm$0.03 &  -0.65$\pm$0.06 \\
6 & 14 15 57.21 & 23 02 58.1 & 1.4 & 22.20$\pm$0.03 & 0.65$\pm$0.05 & -1.51$\pm$0.04 &  -0.32$\pm$0.09 \\
\enddata
\tablenotetext{a}{Concentration Index determined from the difference between the magnitude in an aperture of three pixel
radius compared to that of a one pixel radius in the F606W filter.}
\end{deluxetable}

\begin{deluxetable}{lccccc}
\tabletypesize{\scriptsize}
\tablecolumns{6}
\tablewidth{280pt}
\tablecaption{SED Fits to the Clusters in DDO 187 \label{tab-d187sed}}
\tablehead{
\colhead{ID} & \colhead{Chi-square} & \colhead{log Age (yr)} & \colhead{Z\tablenotemark{a}} & \colhead{A$_V$} & \colhead{log Mass (M\solar)}
}
\startdata    
1 & 1.9 &  $8.8_{0.1}^{0.2}$ & 0.005 & $0.0$\tablenotemark{b} & $3.6_{0.1}^{0.2}$ \\
2 & 2.0 &  $8.7_{0.1}^{0.3}$ & 0.006 & $0.0$\tablenotemark{b} & $3.3_{0.1}^{0.2}$ \\
3 & 2.0 &  $6.5_{0.1}^{0.4}$ & 0.005 & $0.9_{0.6}^{0.6}$           & $2.2_{0.5}^{0.3}$ \\
4 & 0.1 &  $6.9_{0.3}^{0.4}$ & 0.002 & $0.3_{0.3}^{0.4}$           & $2.4_{0.6}^{0.4}$ \\
5 & 2.0 &  $6.5_{0.1}^{0.1}$ & 0.005 & $1.1_{0.4}^{0.5}$           & $2.7_{0.3}^{0.3}$ \\
6 & 2.0 &  $6.5_{0.1}^{0.2}$ & 0.005 & $0.7_{0.6}^{0.5}$           & $2.2_{0.4}^{0.3}$ \\
\enddata
\tablenotetext{a}{Metallicity Z is a free parameter. The oxygen abundance given in Table \ref{tab-sample} corresponds to a Z of about 0.001.}
\tablenotetext{b}{These extinction values A$_V$ were set to 0 in order to better constrain the other parameters.}
\end{deluxetable}

\subsection{Dividing the galaxies into halves}

This study was motivated by the distribution of UV light, and hence young regions, in
DDO 187 and NGC 3738, and we wish to compare the two halves of these regions of the galaxies.
In Figures \ref{fig-d187} and \ref{fig-n3738} we show how we have divided each galaxy in half: the high star-forming
half and the low star-forming half. 
The images in the upper left are the {\it GALEX} FUV-band image, which we use in order to maximize the contrast for young stars.
The white or black line marks the separation between the two halves, determined by eye. 
The bright half of each galaxy (SW for DDO 187 and NW for NGC 3738) 
is referred to as the high star formation (HSF) half and the fainter half 
(NE for DDO 187 and SE for NGC 3738) is referred to as the low star formation (LSF) half. 


We filled the regions 
with circles that together mostly cover the area under study.
The purpose of the circles is to measure the photometric properties of each half of the galaxy.
The photometry on images in the different filters and maps of different environmental properties 
(pressure, \HI\ surface density, stellar mass surface density)
within the circles taken together in each half define the properties in that half of the galaxy.
Red circles outline regions in the HSF half, and green circles outline regions in the LSF half.

The combined areas of the circles in each half of each galaxy are given in Table \ref{tab-galprop}.
The total area in all the circles representing the properties in DDO 187 is 2.1 times the area within one
disk scale length.
The total area in all the circles in NGC 3738, which is much more centrally concentrated than DDO 187,
is 0.1 times the area within one disk scale length.

\section{Results} \label{sec-results}

\subsection{Galactic properties} \label{sec-prop}

\begin{figure}[t!]
\epsscale{0.65}
\plotone{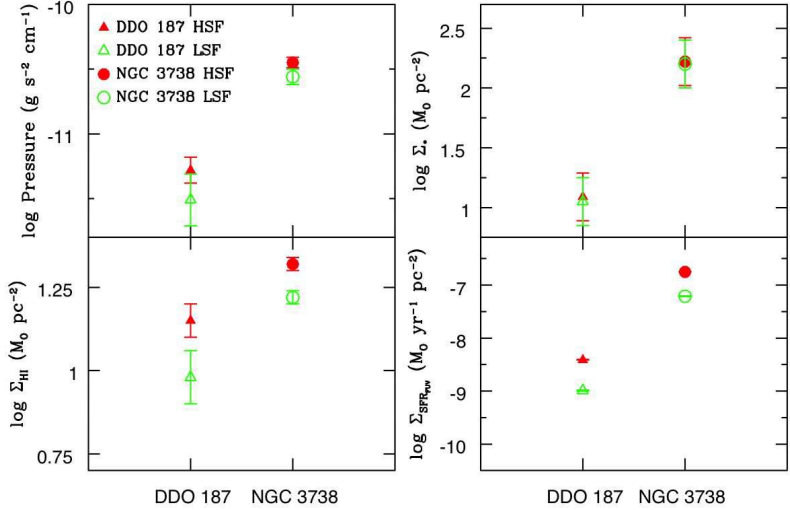}
\vskip -0.1truein
\caption{Pressure, \HI\ mass surface density $\Sigma_{HI}$, stellar mass surface density $\Sigma_{*}$, and
SFR determined from the FUV per unit area $\Sigma_{FUV}$ plotted for the
HSF and LSF regions in DDO 187 and NGC 3738.
}
\label{fig-regprops}
\end{figure}

The physical properties (pressures, stellar mass surface densities, and \HI\ mass surface densities)
in the HSF and LSF regions of DDO 187 and NGC 3738 are given
in Table \ref{tab-galprop} and shown pictorially in Figure \ref{fig-regprops}.
The stellar mass density is about the same in the HSF and in the LSF regions, but the pressures and \HI\ mass surface densities
are higher in the HSF region than in the LSF in both galaxies:
pressure is 30-70\% higher and gas density is 30-50\% higher, with DDO 187 having the more extreme difference between the 
HSF and LSF regions of the two galaxies.

We also give the SFR determined from the {\it GALEX} FUV flux in Table \ref{tab-galprop}.
The FUV averages over the past 100 Myr.
We find that the SFR per unit area is higher in the HSF regions than in the LSF regions by a factor of $\sim$3-4.
These results suggest that either or both of the pressure and \HI\ mass surface density are important in enhancing the SFR in the HSF regions of these galaxies.

\begin{deluxetable}{lccccccc}
\tabletypesize{\tiny}
\tablecolumns{8}
\tablewidth{410pt}
\tablecaption{Galactic Properties\tablenotemark{a} \label{tab-galprop}}
\tablehead{
\colhead{} & \colhead{} & \colhead{Area} & \colhead{Effective radius} & \colhead{$\log \Sigma_{HI}$} & \colhead{$\log {\rm pressure}$} & \colhead{$\log \Sigma_*$} & \colhead{$\log {\rm SFR}_{FUV}$\tablenotemark{b}} \\
\colhead{Galaxy} & \colhead{Region} & \colhead{(kpc$^{2}$)} & \colhead{(pc)} & \colhead{(M\solar pc$^{-2}$)} & \colhead{(g s$^{-2}$ cm$^{-1}$)} & \colhead{(M\solar pc$^{-2}$)} & \colhead{(M\solar yr$^{-1}$ kpc$^{-2}$)}
}
\startdata
DDO 187    & HSF & 0.113 
                   & 190 & 1.15$\pm$0.05  & -11.28$\pm$0.1  & 1.09$\pm$0.2 & -2.41$\pm$0.005 \\
                   & LSF  & 0.108 
                   &  185 & 0.98$\pm$0.08  &  -11.51$\pm$0.2 & 1.05$\pm$0.2 &  -2.99$\pm$0.011 \\
                   & HSF/LSF & \nodata & \nodata & 1.5$\pm$0.3  & 1.7$\pm$0.6 & 1.1$\pm$0.7 & 3.8$\pm$0.11 \\
NGC 3738 & HSF &   0.086 
                  & 165 & 1.32$\pm$0.02 & -10.45$\pm$0.04 & 2.22$\pm$0.2 & -0.75$\pm$0.003 \\
                    & LSF  & 0.117 
                    & 193 & 1.22$\pm$0.02 & -10.56$\pm$0.06 & 2.20$\pm$0.2 & -1.21$\pm$0.006  \\
                    & HSF/LSF & \nodata & \nodata & 1.3$\pm$0.08 & 1.3$\pm$0.2 & 1.0$\pm$0.7 & 2.9$\pm$0.03 \\
\enddata
\tablenotetext{a}{The third line for each galaxy is the ratio of the property in the HSF region to that in the LSF region.}
\tablenotetext{b}{$\log SFR_{FUV}$ is the SFR in each region derived from the FUV luminosity in that region divided by the area of the region given in column 3. See \citet{ludka10} for the formula used for determining the SFR from the FUV luminosity.
}
\end{deluxetable}

\subsection{SFR and gas density}

\begin{figure}[t!]
\epsscale{0.6} 
\vskip -0.5truein
\plotone{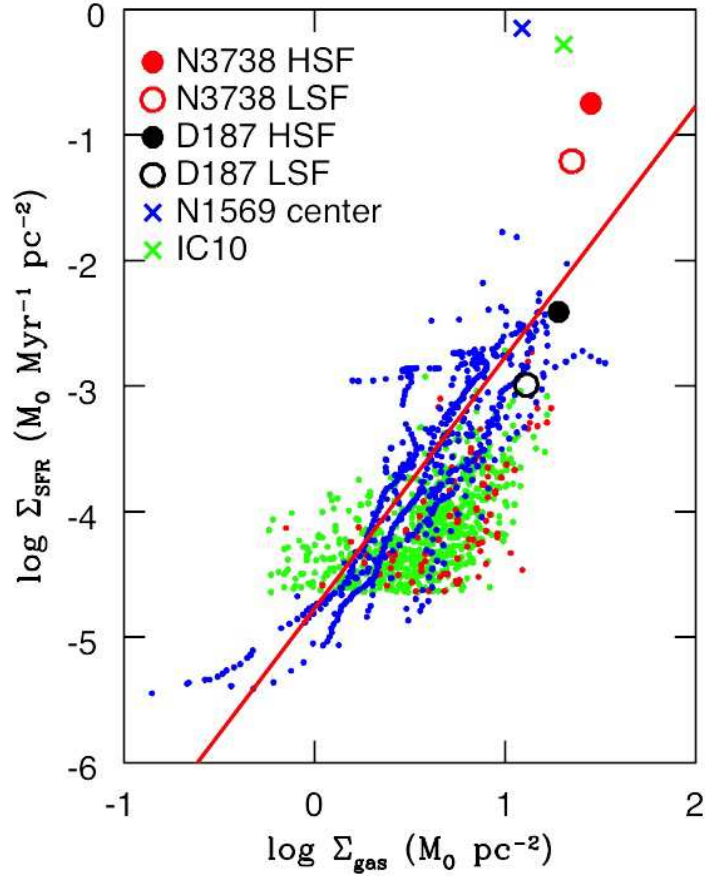}
\vskip -0.25truein
\caption{Kennicutt-Schmidt relation of SFR surface density vs.\ gas surface density for azimuthally averaged annuli and sub-kiloparsec sized regions within
galaxies. The plot is adopted from Figure 1 of \citet{bruce15}, with the SFRs for
the LSF and HSF regions of NGC 3738 and DDO 187 added.
The \HI\ has been corrected for the addition of He and heavy elements but not molecules.
The small green points are THINGS spirals from \citet{bigiel10}, the red points are THINGS dwarfs from \citet{bigiel10}, and
the blue points are LITTLE THINGS dwarfs from \citet{eh15}. The red slanted line is the theoretical prediction from \citet{bruce15}.
The SFR in the HSF region in NGC 3738 is about 13 times higher than predicted by its gas density.
For comparison, we plot the central  208 pc radius of the starburst dIrr NGC 1569
and the big star-forming complex (radius of 196 pc) in IC 10, both of which lie a bit further from the Kennicutt-Schmidt relation 
along the SFR axis than the HSF region in NGC 3738. 
The SFRs are determined from {\it GALEX} FUV fluxes and so integrate over the past 100 Myr for all galaxies, including
comparison surveys, except IC 10.
The SFR of the region in IC 10 is derived from \ha\ and so integrates over the past 10 Myr.
}
\label{fig-KS}
\end{figure}

Empirically there is a relationship between gas density and SFR density in galaxies, known as the Kennicutt-Schmidt relation.
Originally it was noticed for galaxy-wide integrated SFRs and gas densities \citep{kennicutt89,kennicutt98},
but has also been studied in sub-kiloparsec sized
regions within galaxies \citep{bigiel08,bigiel10,ficut}. In the inner parts of spiral galaxies $\Sigma_{SFR}\sim\Sigma_{gas}^{1.4}$, and in the
outer disks of spirals and in dIrr galaxies, the relationship steepens.
In Figure \ref{fig-KS} we use the quantities in Table \ref{tab-galprop} to place the LSF and HSF halves of NGC 3738 and DDO 187 
on a plot of SFR density against gas surface density $\Sigma_{SFR}$ vs.\ $\Sigma_{gas}$. 
The plot is from Figure 1 of \citet{bruce15}. 
The other points on the plot are regions within THINGS spirals and dwarfs from \citet{bigiel10} and
LITTLE THINGS dwarfs from \citet{eh15}. The red line is the theoretical prediction of \citet{bruce15}, included as a model that is
applicable to dIrrs \citep[see also][]{ostriker10,krumholz12}.
We see that the SFR$_{FUV}$ in the HSF region in NGC 3738 is about {\bf 13} times higher than predicted by its gas density.
The LSF region is high too.
On the other hand, 
both of the DDO 187 regions fall within the scatter of points around the theoretical line.
Falling far from the Kennicutt-Schmidt relation in primarily atomic gas is also seen in the interacting galaxies IC
2163 and NGC 2207 by \citet{bruce16}, primarily a result of high compressive turbulence broadening the density probability
density function of the \HI.  The HSF region of NGC 3738 does not have high turbulence itself, but the velocity field does
have peculiar motions overall.

However, the DDO 187 and NGC 3738 regions are not arbitrary regions within the galaxies;
the HSF regions were chosen for being extreme areas of star formation. In addition, the DDO 187 and NGC 3738 regions
are not as large as the areas in the comparison objects in the figure, which are azimuthal averages in annuli or 400 pc to 750 pc size cells.
Larger regions include a more heterogeneous star formation history.  
For comparison we also plot  on Figure \ref{fig-KS} the central 208 pc radius of the dIrr starburst NGC 1569 and the big star-forming complex in IC 10 with
a radius of 196 pc \citep{ludka10,lt,zhang12}, as examples of other intense star-forming complexes. 
The SFR for NGC 1569 is determined from the FUV and hence averages over the past 100 Myr. 
For IC 10 the SFR is from H$\alpha$ and so averages over the past 10 Myr.  Both SFRs are corrected for reddening.
These regions are comparable
in size to the HSF region in NGC 3738. However, in terms of the SFR, the center of NGC 1569 and the region in IC 10 are a bit more
extreme and fall further from the Kennicutt-Schmidt relation.
Thus, the HSF region of NGC 3738 is not unusual in its placement on the Kennicutt-Schmidt relation for a region of intense star formation.

\subsection{Star clusters} \label{sec-cl}
 
\begin{figure}[t!]
\epsscale{0.7} 
\vskip -0.25truein
\plotone{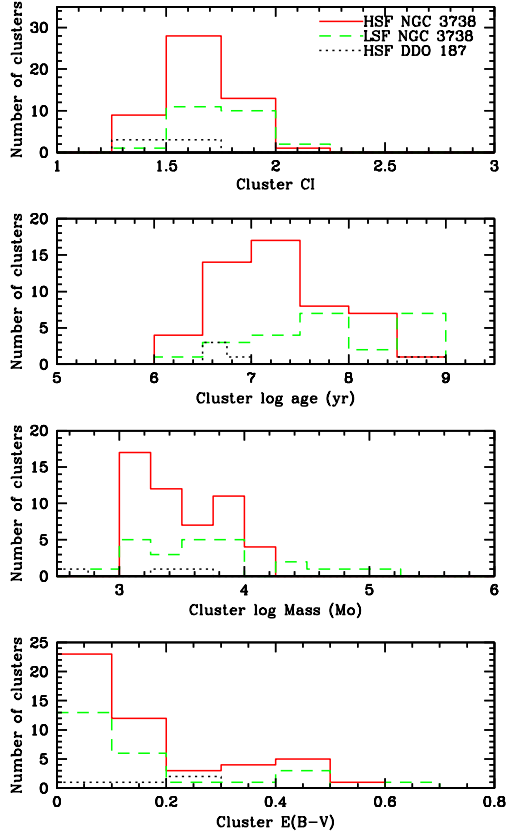}
\caption{Star clusters in the HSF and LSF regions of NGC 3738 and the HSF side of DDO 187.
Clusters were identified and their properties were measured on LEGUS {\it HST} images \citep{legus,leguscl}
for NGC 3738 and on archive {\it HST} images for DDO 187.
In NGC 3738 cluster CI, masses, and E($B-V$) are similar in both halves of the galaxy, but the bulk
of the clusters in the HSF region are younger than in the LSF region.
DDO 187 only has 6 clusters and they are all on the HSF side of the galaxy.}
\label{fig-n3738cl}
\end{figure}

Properties of the compact star clusters identified from {\it HST} images in NGC 3738 are shown for
the two halves of the galaxy in Figure \ref{fig-n3738cl}.
The cluster CI have similar distributions in the HSF and LSF parts of NGC 3738. 
The cluster masses are nearly similar too except that the LSF region has a tail to higher cluster masses that are missing
in the HSF region. 
In addition, the cluster extinction E($B-V$), one of the fit parameters to the photometry, cover the same range in both regions.
On the other hand, the ages are skewed to younger ages in the HSF region.
The bulk of the clusters in the HSF half have ages of 10-20 Myr with the median age being 20 Myr.
The two age peaks in the LSF region
are at 60 Myr and 600 Myr with a median age of 100 Myr.
The HSF half also contains about twice as many clusters as the LSF half: 51 in the HSF half and 24 in the LSF half.

We have performed the Anderson-Darling nonparametric 2-sample test 
\citep{AD,SS}
on the properties of the clusters: 51 clusters in the HSF region and
24 clusters in the LSF region. The resulting p-values for CI, age, and mass are 0.025, 0.002, and 0.15, respectively. 
This implies that the CI values and age are not drawn from the same distribution, but the mass may be.

Because the HSF half is more crowded with clusters and OB stars and the background surface brightness is higher,
incompleteness of the cluster catalog could be higher on the HSF side. This would bias the catalog against fainter,
more diffuse, and older clusters in that region. However, this should not affect the disparity in age distributions
of clusters in the two regions since young clusters in the LSF region are unlikely to have been missed. 
In addition, the bias would not likely be in the sense of missing high mass compact clusters in the HSF region.
Thus, we conclude that incompleteness effects are not likely to operate in the directions necessary to 
reverse these differences between the HSF and LSF sides.

DDO 187 only has 6 clusters and all are found in the HSF side of the galaxy.
They are compact, four are younger than 10 Myr and two are over 100 Myr in age, masses are
100 M\solar\ to $10^4$ M\solar, and reddenings are less than E($B-V$) of 0.3.

\subsection{Gas kinematics} \label{sec-gas}

In Figures \ref{fig-d187} and  \ref{fig-n3738} we show the integrated \HI\ maps of DDO 187 and NGC 3738,
the intensity-weighted \HI\ velocity fields,
and the intensity-weighted velocity dispersion maps.
We see that both DDO 187 and NGC 3738 show overall rotation but with significant peculiar features as well.
In NGC 3738, the central region of the galaxy is located at a bend in a backward S-shaped iso-velocity contour.

The velocity dispersion maps reveal a peculiarity common to the two galaxies. 
A large region of high velocity dispersion is located just off the regions under discussion here.
In DDO 187 the region is comparable in size to our region and is located
to the north of the LSF region with dispersions of order 14-20 \kms.
In NGC 3738 the high velocity dispersion region has dispersions of 24-35 \kms, is located to the east and north of the optical
center of the galaxy, and is much larger than our target region.
A typical non-starburst dIrr would have a peak velocity dispersion of order 10 \kms, so these regions in
DDO 187 and NGC 3738 have higher velocity dispersions.
In both systems there is also a smaller region of high velocity dispersion on the opposite side of the galaxy.
There is no direct relationship between these high velocity dispersion regions and the HSF regions of the galaxies.
The high velocity dispersion gas is unlikely to be due to energy injected by supernova remnants since
none have been found in NGC 3738 \citep{n3738-sn}.

\begin{figure}[t!]
\epsscale{0.6} 
\includegraphics{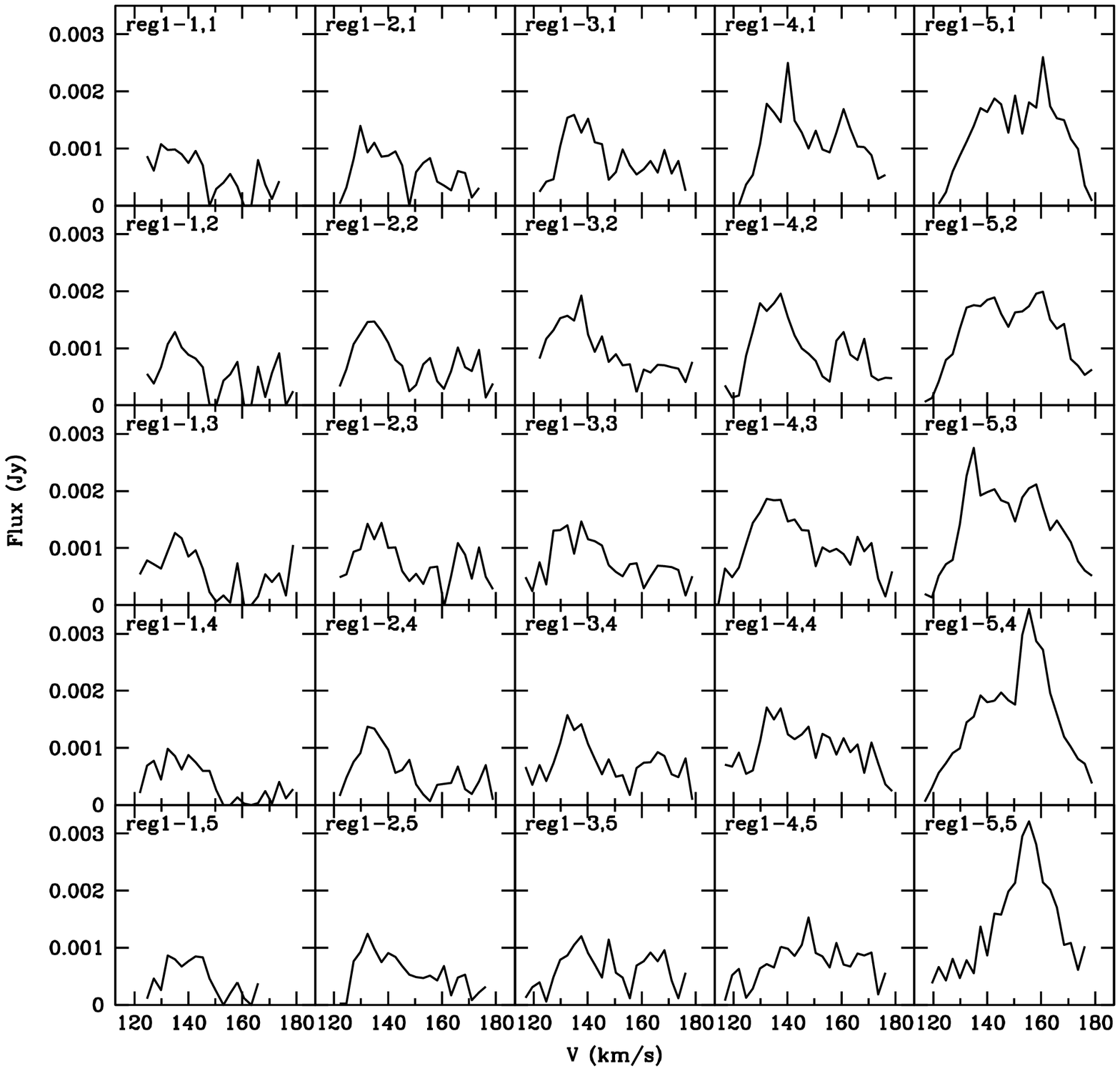}
\vskip 5.0truein
\caption{Velocity profiles at a grid of points over the NE high velocity dispersion area in DDO 187.
The grid is centered at at an RA, Dec of 14h15m57.36s, 23d 4m 0.9s (see Figure \ref{fig-d187}).
A 52\farcs5$\times$52\farcs5 box is covered by a 5$\times$5 grid, where each point averages over
10\farcs5, approximately the beam size of the \HI\ data cube.
The grid number is noted in the upper left of each panel, where position 1,1 is in the NE (top left) part of the
grid and position 5,5 is in the SW (bottom right) part of the grid.
}
\label{fig-d187reg1profs}
\end{figure}

\begin{figure}[t!]
\epsscale{0.6} 
\includegraphics{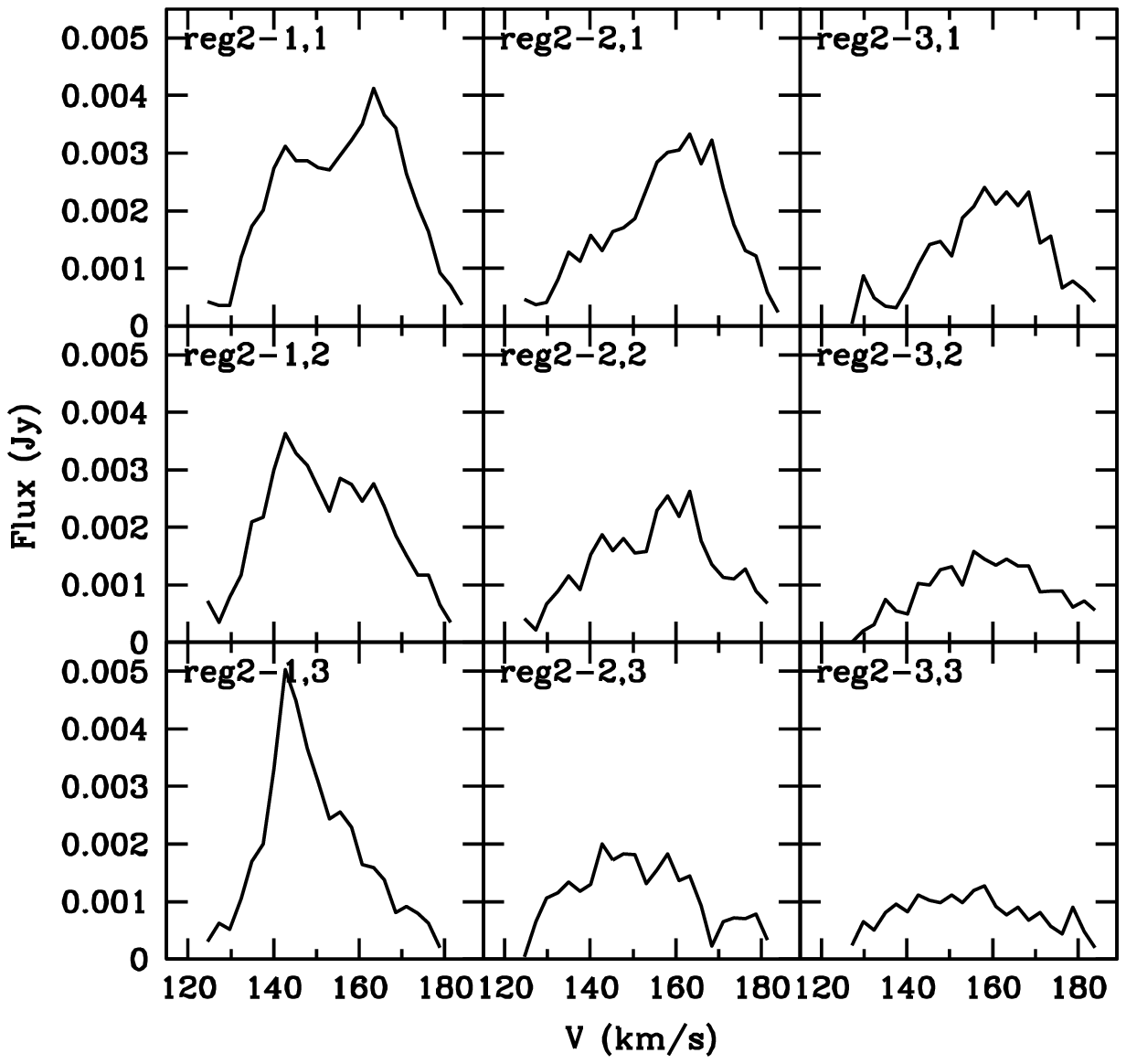}
\vskip 3.0truein
\caption{Velocity profiles at a grid of points over the SW high velocity dispersion area in DDO 187.
The grid is centered at an RA, Dec of 14h15m55.63s, 23d 2m 48.9s (see Figure \ref{fig-d187}).
A 31\farcs5$\times$31\farcs5 box is covered by a 3$\times$3 grid, where each point averages over
10\farcs5, approximately the beam size of the \HI\ data cube.
The grid number is noted in the upper left of each panel, where position 1,1 is in the NE (top left) part of the
grid and position 3,3 is in the SW (bottom right) part of the grid.
}
\label{fig-d187reg2profs}
\end{figure}

To examine the nature of the velocity dispersions further, we plot velocity profiles at points in a rectangular grid
in the high velocity dispersion regions.
In each galaxy there are two or three of these areas, and we sum each point in the grid over a region approximately the size of the \HI\ data cube beam.
For DDO 187 one region is to the NE and one to the SW of the galaxy center and the grid boxes are outlined in Figure \ref{fig-d187}. 
The profiles for DDO 187 are shown in Figures \ref{fig-d187reg1profs} and \ref{fig-d187reg2profs}.
Although the spectra are noisy, in both regions we see complex profiles.
Some profiles could be fit with a single fat gaussian, and others would require multiple gaussians.
So it is not simple turbulence (wide FWHM of the profiles) that is producing these high velocity dispersion regions in the moment 2
map of DDO 187, but also rather complex motions of the gas.

\begin{figure}[t!]
\epsscale{0.6} 
\includegraphics{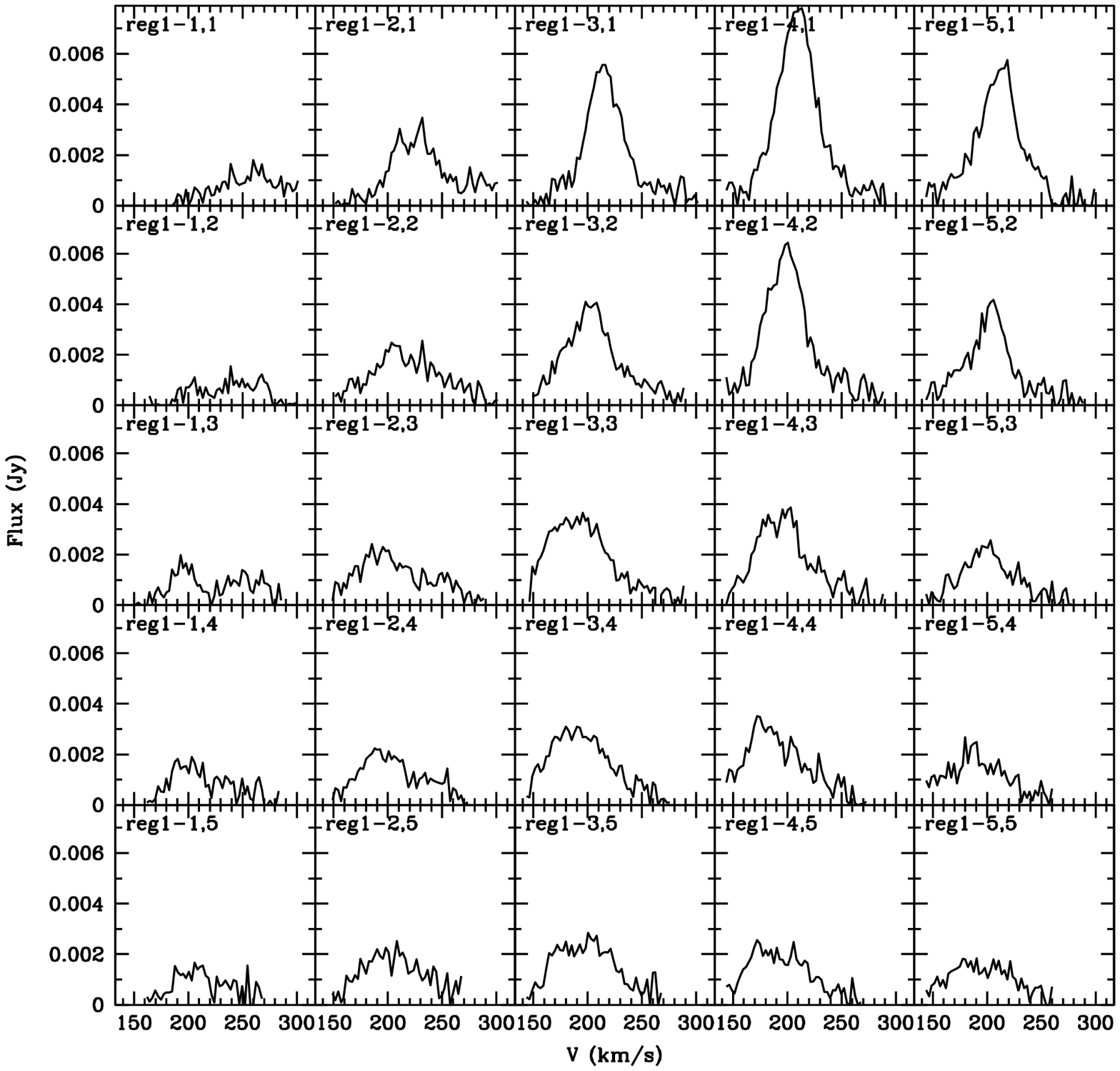}
\vskip 5.0truein
\caption{Velocity profiles at a grid of points over the NE high velocity dispersion area in NGC 3738.
The grid is centered at at an RA, Dec of 11h35m51.90s, 54d31m 51.4s, and outlined in Figure \ref{fig-n3738}.
A 60\arcsec$\times$60\arcsec\ box is covered by a 5$\times$5 grid, where each point averages over
12\arcsec, approximately the beam size of the \HI\ data cube.
The grid number is noted in the upper left of each panel, where position 1,1 is in the NE (top left) part of the
grid and position 5,5 is in the SW (bottom right) part of the grid.
}
\label{fig-n3738reg1profs}
\end{figure}

\begin{figure}[t!]
\epsscale{0.6} 
\includegraphics{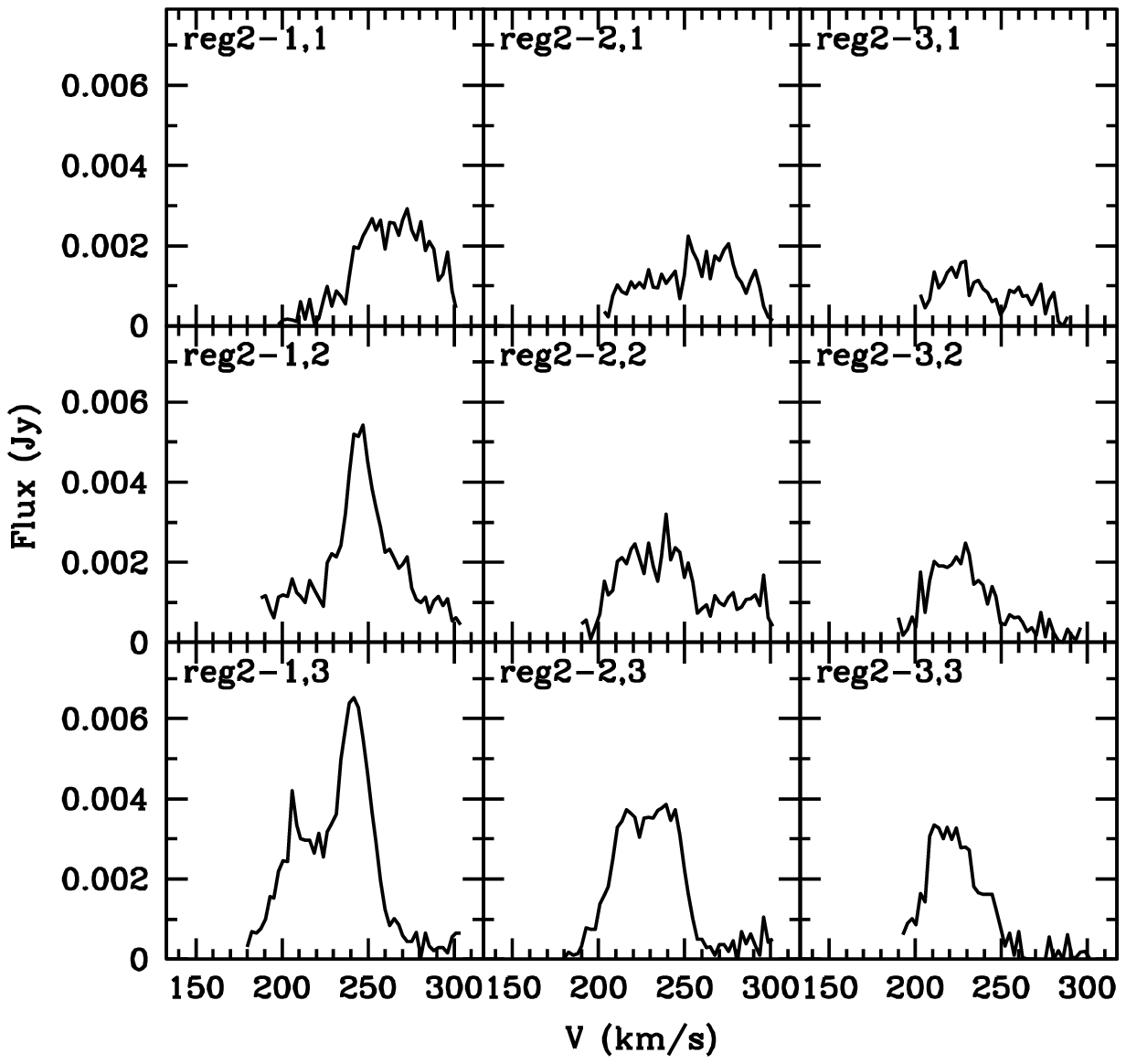}
\vskip 2.95truein
\caption{Velocity profiles at a grid of points over the SW high velocity dispersion area in NGC 3738.
The grid is centered at at an RA, Dec of 11h35m45.00s, 54d31m 21.4s, and outlined in Figure \ref{fig-n3738}.
A 36\arcsec$\times$36\arcsec\ box is covered by a 3$\times$3 grid, where each point averages over
12\arcsec, approximately the beam size of the \HI\ data cube.
The grid number is noted in the upper left of each panel, where position 1,1 is in the NE (top left) part of the
grid and position 3,3 is in the SW (bottom  right) part of the grid.
}
\label{fig-n3738reg2profs}
\end{figure}

\begin{figure}[t!]
\epsscale{0.6} 
\includegraphics{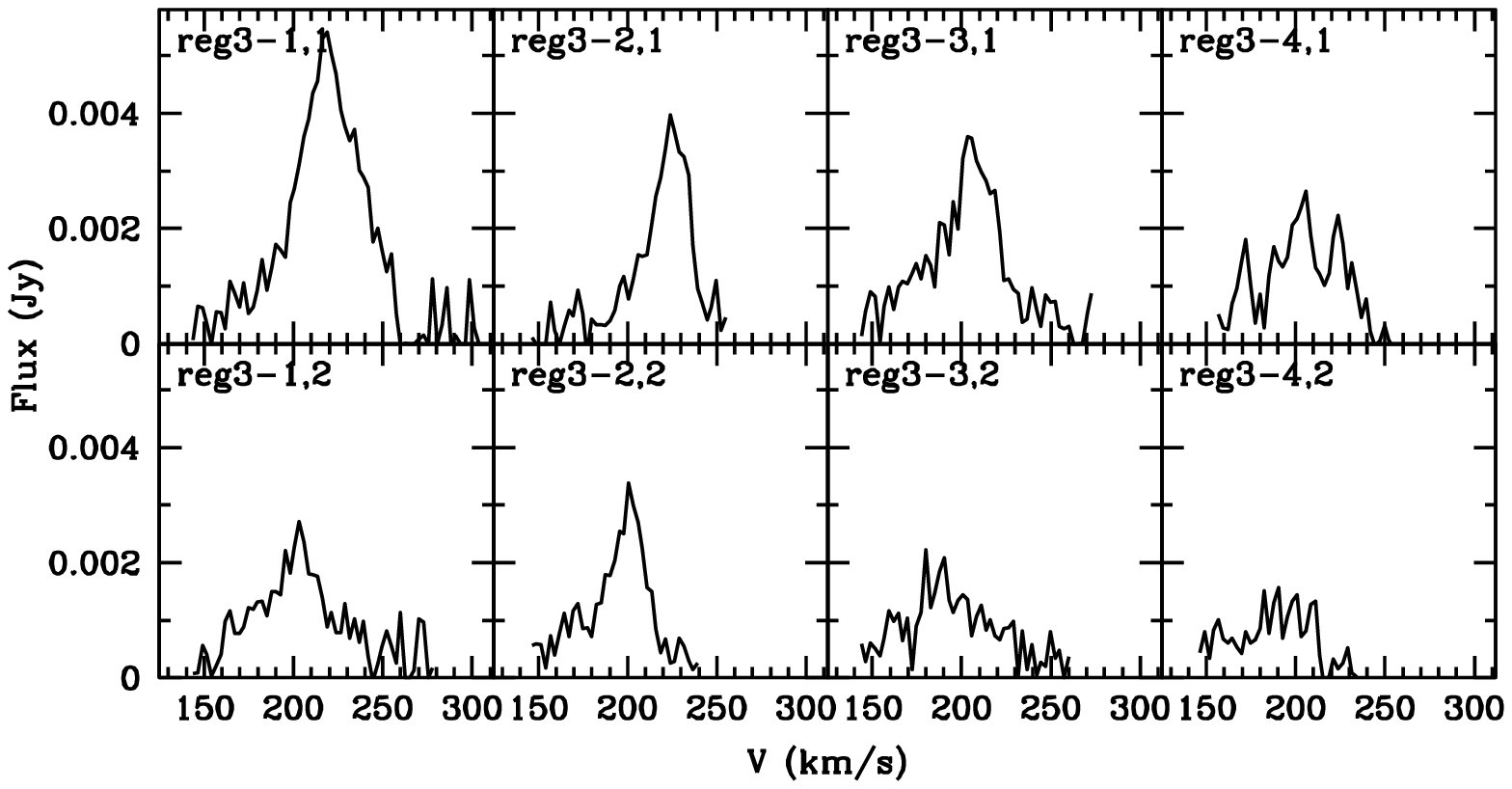}
\vskip 2.2truein
\caption{Velocity profiles at a grid of points over the central high velocity dispersion area in NGC 3738.
The grid is centered at at an RA, Dec of 11h35m50.66s, +54d31m18.5s, and outlined in Figure \ref{fig-n3738}.
A 48\arcsec$\times$24\arcsec\ box is covered by a 4$\times$2 grid, where each point averages over
12\arcsec, approximately the beam size of the \HI\ data cube.
The grid number is noted in the upper left of each panel, where position 1,1 is in the NE (top left) part of the
grid and position 4,2 is in the SW (bottom  right) part of the grid.
}
\label{fig-n3738reg3profs}
\end{figure}

The velocity profile grids of the two major high velocity dispersion areas and one in between in NGC 3738 are shown in 
Figures \ref{fig-n3738reg1profs}, \ref{fig-n3738reg2profs}, and \ref{fig-n3738reg3profs}, 
and outlined with boxes on the velocity dispersion map in Figure \ref{fig-n3738}.
As with DDO 187, we see some
profiles that are broad and some that are multiple gaussians.
In fact, 
\citet{oh15} isolate \HI\ gas that is not engaged in ordered rotation in NGC 3738,
and  the integrated \HI\ of the strong non-circular motion gas in NGC 3738
is shown as black contours on Figure \ref{fig-n3738}.
``Strong'' refers to an \HI\ component that has a higher intensity than the gas engaged in ordered
motion at that position in the galaxy. 
The map shows that gas with peculiar motions is located in the high velocity dispersion regions.
The contour in the NE region appears in the lower left quadrant of the grid in Figure \ref{fig-n3738reg1profs}
and the contour in the SW region is the center part of the grid in Figure \ref{fig-n3738reg2profs}.

\section{Discussion} \label{sec-discuss}

We initiated the study of DDO 187 and NGC 3738 because of the similar and unusual morphology 
of the star-formation activity in these galaxies.
The areas whose star-forming properties we have examined were chosen from the optical and NUV appearance. 
The $V$-band isophote used as a boundary in DDO 187 has a radius of about 220 pc and in NGC 3738 a radius of about 330 pc
\citep{HE06}.
These scales are larger by factors of 5-6 than typical OB associations that generally have radii of 50 pc \citep[for example,][]{obassocsize}, 
but are comparable to some star-forming {\it complexes} in dIrrs.
Nevertheless, they represent a significant portion of the optical component of the inner galaxy.
In DDO 187 and NGC 3738 we noticed a stark division of these regions into a high SFR half and a lower SFR half, and undertook
an examination of the properties within the galaxies that might influence this star formation distribution.

Although the SFR in the HSF region of NGC 3738 is 46$\pm$0.6 times higher than that in DDO 187, the two galaxies
have similarities.
Our examination of the HSF and LSF halves of the regions of DDO 187 and NGC 3738 show that
1) the SFR on the HSF side has been 3-4 times higher than that on the LSF side over the past 100 Myr,
2) the pressure and gas density are higher on the HSF sides by 30-70\%,
3) the \HI\ velocity fields exhibit significant deviations from ordered rotation,
and 4) there are large areas of high velocity dispersion and complex motions of the gas in the outer galaxies,
including a gas component with peculiar motions in the high velocity dispersion region of NGC 3738.
We also find that the HSF region in NGC 3738 has a SFR that is 13 times higher than the rate predicted by
its gas density from the Kennicutt-Schmidt relationship, which is similar to other intense star forming complexes
in starburst dwarfs.
The compact star clusters in NGC 3738 are generally younger on the HSF side - 20 Myr median age versus 100 Myr,
and the six clusters in DDO 187 are all located on the HSF side of the galaxy.

The size of the HSF region in NGC 3738 is comparable to the gas scale-height, estimated to be 340 pc at a radius of 0.7 kpc \citep{eh15}.
This could argue that the HSF region represents the limit of what one might expect for a star-forming complex in this galaxy,
and therefore, perhaps not remarkable at all. However, there are other characteristics of the galaxy that suggest
that more is going on.
NGC 3738 and two other LITTLE THINGS BCDs are the subject of a study by \citet{ashley17} in which they examine the
\HI\ distribution and kinematics from the VLA data and look for faint, low column density extended gas at 21-cm with
the highly sensitive Green Bank Telescope\footnote[26]{
A facility of the Green Bank Observatory, which is a facility of the National Science Foundation and is operated by Associated Universities, Inc.}
(GBT). 
In spite of the kinematic and morphological peculiarities in the \HI, NGC 3738 shows no evidence of extended
tails that would be the hallmark of an interaction with another galaxy. 
On the other hand, NGC 3738 does have a kinematically distinct cloud located SW of the galaxy center 
\citep[see Figures 21 and 23 in][]{ashley17} and clumps of low surface density \HI\  in the extended gas \citep{lt}.

\citet{ashley17} suggest three possibilities for the state that NGC 3738 is in: 1) an advanced merger, 2) ram pressure stripping,
or 3) an outflow created by the starburst.
\citet{ashley17} argue that for ram pressure stripping by the Canes Venatici I intergroup medium to be the explanation
NGC 3738 would have to be moving through it face-on to create the azimuthally symmetrical depletion of extended gas.
An outflow due to the starburst could puff up the stellar and gaseous components of the galaxy, according to \citet{elbadry16},
and perhaps account for the large size of the stellar disk relative to the \HI\ and for the scattered, low column density clouds
to the SW and North of the main galaxy (see Figure \ref{fig-n3738} here and Figure 72 in Hunter et al.\ 2012).
An advanced merger, on the other hand, could also account for the central concentration of gas, as gas is funneled towards
the center in a merger, and the peculiar cloud to the SW
of center, as well as the scattered faint clouds. 


There is a third galaxy in the LITTLE THINGS sample that bears some resemblance to NGC 3738 and DDO 187: IC 10.
In $U$-band the galaxy is dominated by a large UV-bright region near the center of the stellar
component of the galaxy
with two stellar wind-blown bubbles from an older generation of stars to the NW.
The \HI\ 
is highly peculiar: the velocity field is chaotic, the velocity
dispersion is $>20$ \kms\ over much of the gas, and there is a large fan of material extending to the South of the optical 
galaxy. 
Unlike DDO 187 and NGC 3738, however, IC 10 has been found to have an extended \HI\ tail topped with a higher column density
blob of gas, suggesting that IC 10 is an advanced merger or interaction
\citep{ic101,ic102}. 
Star formation in DDO 187 and NGC 3738 is more sedate than that in IC 10, and so if DDO 187 and NGC 3738 are also the results of mergers, 
they must be 
further advanced and/or the parameters of the interaction, such as mass ratios, must have been such as to result in a gentler interaction.

DDO 187 does potentially have two ultra-faint dIrrs nearby that were identified originally as high velocity clouds
\citep{adams16}.
One of these objects, AGC 249525, has an optical counterpart with a red giant branch that suggests a distance
of 1.64$\pm$0.45 Mpc and a baryonic mass of order $10^6$ M\solar\ \citep{janesh17}.  It also has a ``head-tail'' structure 
possibly indicative of a past interaction or ram pressure stripping.
At 1.64 Mpc its distance from DDO 187 would be 680 kpc.
Thus, AGC 249525 is not close enough to have interacted recently. 

\begin{figure}[t!]
\epsscale{0.8} 
\plotone{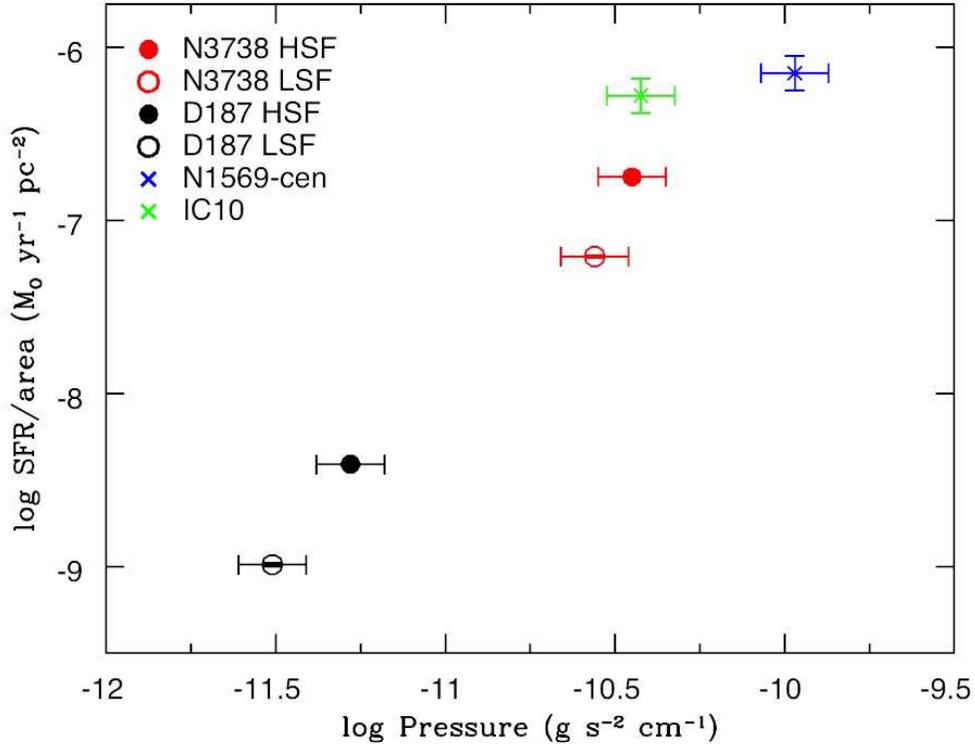}
\vskip -0.1truein
\caption{Pressure vs.\ SFR surface density in the HSF and LSF regions of DDO 187 and NGC 3738,
the large star forming region of IC 10, 
and the center of NGC 1569.
NGC 1569 and IC 10 are both starbursts likely triggered by an external disturbance. 
The SFR uncertainties for NGC 3738 and DDO 187 are smaller than the symbols.
}
\label{fig-presvssfr}
\end{figure}

In order to compare the properties of the region in IC 10 with those in NGC 3738 and DDO 187,
we plot pressure against the SFR per unit area in Figure \ref{fig-presvssfr}.
The center of NGC 1569 is included in this comparison. 
Both NGC 1569 and IC 10 are starbursts with the likely explanation being a gravitational interaction with another galaxy.
The region over which we
measured the properties in IC 10 has a radius of 196 pc. 
There is no {\it GALEX} FUV image for IC 10,
so the SFR is determined from the \ha\ image using the relationship given in \citet{ludka10} that is based
on that given by \citet{kennicutt98}.
We see that the SFR per unit area $\Sigma_{\rm SFR}$ increases with pressure, and the HSF region in NGC 3738 is similar in
properties to the region in IC 10. The center of NGC 1569 is the most extreme in this plot, having a higher pressure and $\Sigma_{\rm SFR}$.

There is also a similarity between the star formation distributions found in DDO 187 and NGC 3738
and those seen in ``tadpole'' galaxies in the sense that these systems are dominated by a
giant star-forming region. 
In the case of tadpole galaxies the giant star-forming regions are located at the end of an elongated distribution of stars, the head of the tadpole
\citep{tadpolehdf,tadpolelocal}. The head of tadpole galaxies found in the Hubble Deep Field have 
higher masses of stars by factors of 100-1000 than we find in the HSF regions of DDO 187 or NGC 3738. 
However, local tadpoles have been found that are comparable to DDO 187 and NGC 3738, and
this morphology is relatively common among extremely metal-poor galaxies \citep{xmp}.
Furthermore, at least one tadpole head has been found to have a much higher SFR for its probable gas density \citep{tadpoleaccretion2}, 
similar to the HSF region in NGC 3738 as shown in Figure \ref{fig-KS}.
Measurements of the oxygen abundance in several tadpoles revealed that the heads are more metal-poor than 
the bodies \citep{tadpoleaccretion, tadpoleaccretion2}, and
this characteristic of the tadpoles suggests the possibility that the star-forming head is the result of infall of metal-poor gas \citep{tadpoleaccretion2,tadpoleaccretion3}. 
On the other hand, the oxygen abundances of DDO 187 and NGC 3738 are fairly high (see Table \ref{tab-sample}),
which would be inconsistent with accretion of metal-poor gas from the cosmic web.

For DDO 187 and NGC 3738 we can draw the following picture:
the higher pressure and gas densities in the HSF regions have enabled star formation in those regions today. 
Those characteristics in turn resulted from the general chaos in the velocity field, and
for NGC 3738, possibly the impact of a ``rogue'' gas cloud just outside the central region identified by \citet{ashley17} could also have played a role.
In NGC 3738, the LSF half of the galaxy had a higher SFR than the HSF side over 100 Myr ago and 
the massive stars formed in that event could also have created higher density gas in the surroundings
\citep[for example,][]{dopita85,ee98b},
with more spectacular results in terms of high $\Sigma_{gas}$ towards the center of the galaxy where the HSF region is located
today than towards the outer disk. 
The larger scale conditions in these galaxies in turn could have resulted from the merger of two dIrrs.
Unfortunately, causal connections are speculative at this stage of our knowledge,
but the star-forming pattern does suggest some large-scale influence on conditions for star formation in these galaxies.

\section{Summary} \label{sec-summary}

In the dIrr galaxies DDO 187 and NGC 3738 we find that the current star formation is taking place in a large region 
occupying half of the inner optical galaxy. 
We have examined the properties of the high SFR halves of the galaxies 
for comparison with the lower SFR halves
in order to determine what might have influenced this star formation pattern.
DDO 187 is not as extreme as NGC 3738, but there are similarities.
We find the following: 

\begin{enumerate}
\item The SFR averaged over the past 100 Myr on the HSF side is a factor of a few times higher than that on the LSF side in
both galaxies.
\item The compact star clusters on the HSF side of NGC 3738 have a median age of 20 Myr, which is younger than the median age on the LSF side.
\item The pressure and gas density are higher on the HSF sides by 30-70\%, 
implying that the pressures and/or \HI\ mass surface densities are important in enhancing the SFR in the HSF regions
of these galaxies.
\item The SFR averaged over the past 100 Myr on the HSF side of NGC 3738 is higher by a factor of 13 than the rate predicted by its gas density
from the Kennicutt-Schmidt relationship, which is similar to other star forming complexes in starburst dIrrs.
\item The \HI\ velocity fields exhibit significant deviations from ordered rotation.
\item There are large regions of high velocity dispersion and complex kinematics in the gas beyond the central regions of the galaxies.
\end{enumerate}

There are many similarities of DDO 187 and NGC 3738 with the dIrr IC 10, 
a dwarf with a tidal tail that is likely the product
of a merger or interaction of two dwarfs.
We conclude that the higher pressure and gas density in the HSF regions of DDO 187 and NGC 3738 could have 
enabled the current star formation there. These conditions in turn are likely the result of large-scale processes
affecting the internal environment of the galaxies.

\acknowledgments
Some of the results presented here are based on observations made with the NASA/ESA {\it Hubble Space Telescope}
under the LEGUS survey.
Support for Program number 13364 was provided by NASA through a
grant from the Space Telescope Science Institute, which is operated by the
Association of Universities for Research in Astronomy, Incorporated, under NASA
contract NAS5-26555.
AA acknowledges the support of the Swedish Research Council (VetenskapsrŒdet) and the Swedish National Space Board (SNSB).
MF acknowledges support by the UK Science and Technology Facilities Council (grant number ST/P000541/1).
SG appreciates funding from the National Science Foundation grant AST-1461200 to Northern Arizona University for 
Research Experiences for Undergraduates summer internships 
and Drs.\ Kathy Eastwood and David Trilling for running the NAU REU program in 2016.
SG also appreciates the support of the 2016 CAMPARE Scholar program and Dr.\ Alexander Rudolph for directing that program.

Facilities:  \facility{VLA}, \facility{HST(ACS,UVIS)}, \facility{Lowell Observatory}


\end{document}